\documentstyle[12pt]{article}
\begin{document}
\thispagestyle{empty}
\begin{center}
\LARGE\tt\bf{On spin-driven inflation from inflaton fields in General Relativity and COBE data}
\end{center}
\vspace{5cm}
\begin{center}
{\large By L.C. Garcia de Andrade\footnote{Departamento de F\'{\i}sica Te\'{o}rica - Instituto de F\'{\i}sica - UERJ - Rua S\~{a}o Fco. Xavier 524, Rio de Janeiro, RJ, Maracan\~{a}, CEP:20550-003.}}
\end{center}
\begin{abstract}
Obukhov spin-driven inflation in General Relativity is 
extended to include inflaton fields.A de Sitter phase 
solution is obtained and new slow-rolling conditions for 
the spin potential are obtained.The spin potential reduces 
to Obukhov result at the present epoch of the Universe where 
the spin density is low with comparison to the Early 
Universe spin densities.A relation betwenn the spin density energy and the temperature fluctuation can be obtained which allow us to determine the spin density energy in terms of the COBE data for temperature fluctuations.
\end{abstract}
\newpage
Recently I showed \cite{1}that COBE results on temperature 
fluctuations could be used to compute the spin-torsion 
density \cite{2} at the inflation de Sitter phase of the 
Universe and at the present epoch.More recently Ramos and 
myself \cite{3} investigated the role of spin-torsion on the 
chaotic phase of the Universe  showing that the spin and 
torsion interactions are effectively only at the very first 
e-folds of inflation becoming quickly negligible and therefore
not affecting the standard inflationary scenario nor the 
density perturbations.Earlier L.H.Ford \cite{4} proposed a 
model of a vector driven inflation instead of the more usual 
models of scalar inflation proposed by Linde \cite{5} and Guth \cite{6}.
More recently Obukhov \cite{7} extended Ford's work to include a spinning fluid like in torsion theories of gravity building a spin-driven inflation.In this letter we extend Obukhov idea to allow for a double vector-scalar inflation where the role of the scalar field is played by the inflaton and the vector role is played by the spin in the context of general relativity.This work can be further generalized if we also allow for the presence of torsion.But spin-torsion inflation with scalar field have been extensively investigated.Some interesting results appear from our model such as a new slow rolling condition for the spin potential in analogy to the slow rolling condition for the standard inflationary scenario.
At the present epoch of the Universe both spin potentials coincide.Besides we compute the spin density energy from the COBE data of temperature fluctuations and found a result very similar to Obukhov one.Our dynamical equations are easily obtained from the Obukhov ones by the following prescription.We must substitute the spinning fluid pressure and matter densities by the following effective densities
\begin{equation}
e_{eff}={\dot{\phi}}^{2}+V({\phi})
\label{1}
\end{equation}
and
\begin{equation}
p_{eff}={\dot{\phi}}^{2}-V({\phi})
\label{2}
\end{equation}
where ${\phi}$ is the inflaton potential and $V({\phi})$ is the inflaton potential which throughtout this work we shall take to be $V({\phi})=\frac{{m}^{2}{\phi}^{2}}{2}$.Taking the usual Friedmann cosmological metric
\begin{equation}
ds^{2}=dt^{2}-a^{2}(t)(dx^{2}+dy^{2}+dz^{2})
\label{3}
\end{equation}
Obukhov dynamical spinning fluid equations endowed with dilaton fields read
\begin{equation}
3\frac{{\dot{a}}^{2}}{a^{2}}=k(e_{eff}+V_{spin})
\label{4}
\end{equation}
and
\begin{equation}
-(2\frac{\ddot{a}}{a}+\frac{{\dot{a}}^{2}}{a^{2}})=k(p_{eff}-V_{spin}+2{\sigma}^{2}{V'}_{spin})
\label{5}
\end{equation}
where here $ i,j=0,1,2,3 $ , $S_{ij}$ is the spin density 
tensor and ${\sigma}^{2}=S_{ij}S^{ij}$ is the spin density energy.Here the dash represents derivative with respect to ${\sigma}^{2}$.The conservation law is given by
\begin{equation}
{\dot{e_{eff}}}+3{\frac{\dot{a}}{a}}(e_{eff}+p_{eff})=0
\label{6}
\end{equation}
To consider de Sitter inflation we also perform the substitution $\frac{\dot{a}}{a}=H=constant$.From expressions (\ref{1}),(\ref{2}) and (\ref{6}) one obtains 
\begin{equation}
2{\dot{\phi}}{\ddot{\phi}}+{\dot{V}}({\phi})+6H{\dot{\phi}}^{2}=0
\label{7}
\end{equation}
which reduces to
\begin{equation}
2\frac{\ddot{\phi}}{\dot{\phi}}+\frac{\dot{V}}{\dot{\phi}^{2}}=-6H
\label{8}
\end {equation}
By making use of the following approximation $|\frac{\dot{\phi}}{\dot{\phi}}|<<|\frac{\dot{V}}{\dot{\phi}^{2}}|$ into (\ref{8}) yields  
\begin{equation}
{\dot{V({\phi})}}=-6H{\dot{\phi}}^{2}
\label{9}
\end{equation}
which yields
\begin{equation}
{\dot{\phi}}+\frac{m^{2}{\phi}}{6H}=0
\label{10}
\end{equation}
whose solution is
\begin{equation}
{\phi}(t)=e^{-{\frac{m^{2}}{6H}}t}+d
\label{11}
\end{equation}
where $d$ is an integration constant.By equating equations (\ref{4}) and (\ref{5}) above one has
\begin{equation}
{\alpha}^{2}=-\frac{1}{2}(V_{spin}+{\sigma}^{2}{V'}_{spin})
\label{12}
\end{equation}
where ${\alpha}^{2}=[1+\frac{m^{4}}{24H^{2}}]$.From expressions (\ref{12}) one can derive a qualitative slow rolling type condition for $V_{spin}$ as
\begin{equation}
|\frac{{V'}_{spin}}{V_{spin}}|<{\sigma}^{2}
\label{13}
\end{equation}
This is similar to the condition \cite{6} $|\frac{V'}{V}|<<<H^{2}$.
Equation (\ref{12}) can be easily integrated since it reduces to
\begin{equation}
\frac{d{V_{spin}}}{d{{\sigma}^{2}}}={\alpha}^{2}
\label{14}
\end{equation}
Thus equation (\ref{14}) yields an immediate integration to
\begin{equation}
V_{spin}={\alpha}^{2}{\sigma}^{2}+c
\label{15}
\end{equation}
where $c$ is another integration constant.This result is of 
course distinct from Obukhov potential 
\begin{equation}
{V^{O}}_{spin}=(1-e^{-{\beta}{\sigma}^{2}})
\label{16}
\end{equation}
nevertheless at the present epoch of the Universe where the 
spin density is known to be very low \cite{1},it reduces to
\begin{equation}
{V^{O}}_{spin}={\beta}{\sigma}^{2}
\label{17}
\end{equation}
where ${\beta}$ is another constant.Therefore both equations (\ref{15}) and (\ref{17}) 
coincides if $c=0$ and ${\beta}={\alpha}^{2}$.To resume we 
obtain a solution of a mixed spin-inflaton inflation where to 
spin-spin interactions we add the inflaton field.Finally let us determine the spin density energy in terms of the temperature fluctuation ${\frac{{\delta}T}{T}}$.This can be done from equation (\ref{5}) above for the de Sitter phase.After some algebra we obtain
\begin{equation}
k{\sigma}^{2}=\frac{3H^{2}e^{\frac{m^{2}}{3H}}}{\frac{m^{2}}{2}-\frac{m^{4}}{36H^{2}}}
\label{18}
\end{equation}
Since quantum fluctuations \cite{8} are important when $m^{2}<<H^{2}$ we may simplify expression (\ref{18}) to
\begin{equation}
k{\sigma}^{2}=\frac{3H^{2}}{2m^{2}}
\label{19}
\end{equation}
Since the dilaton mass can be written in terms of the Planck mass $m_{Pl}$ as $m=10^{-4}m_{Pl}$ one obtains \cite{8} 
\begin{equation}
k{\sigma}^{2}=\frac{3.10^{8}H^{2}}{2m^{2}}=\frac{3}{2}|{\frac{{\delta}T}{T}}|^{2}
\label{20}
\end{equation}
Since the Einstein gravitational constant is $k=\frac{8{\pi}G}{c^{3}}=10^{-38}$, substitution of this value into equation (\ref{20}) yields
\begin{equation}
{\sigma}^{2}=\frac{3.10^{46}}{2}|{\frac{{\delta}T}{T}}|^{2}
\label{21}
\end{equation}
Since from the COBE data \cite{9} ${\frac{{\delta}T}{T}}=10^{-5}$ and expression (\ref{21}) one obtains
\begin{equation}
{\sigma}^{2}=\frac{3.10^{36}}{2}cgs units
\label{21}
\end{equation}
This result shows that primordial spin density energies can be extremely high as estimated from the COBE data.Obukov obtained 
a value of $10^{40}$ for the spin density by making use of purely spin-driven inflation without inflatons.Our results here can be compared with the results using the Einstein-Cartan equations.Our main conclusion is that as far as these results are concerned we cannot decide between General relativity and Einstein-Cartan theory of gravity.
\vspace{1cm}
\section*{Acknowledgements}
I would like to express my gratitude to Prof.I.Shapiro and 
Prof.H.P. de Oliveira for helpful discussions on the subject 
of this paper.Thanks are also due to CNPq. for partial 
financial support.

\end{document}